\documentclass[prb,
twocolumn,
superscriptaddress,showpacs,amsmath,amssymb]{revtex4}

\usepackage{color}

\begin{document}

\title{Thermodynamics of homogeneous Universes: de Sitter, Bonnor-Melvin and static Einstein }

\author{G.E.~Volovik}
\affiliation{Landau Institute for Theoretical Physics, acad. Semyonov av., 1a, 142432,
Chernogolovka, Russia}


\begin{abstract}
 In the theories, in which dynamic gravitational field emerges from the underlying matter fields, the gravitational field can be considered as a part of matter.  Using this approach, we construct the thermodynamics of the homogeneous Universes -- the de Sitter Universe, the Bonnor-Melvin-$\Lambda$ Universe and the static Einstein Universe. It is demonstrated that although these three Universes have different types of matter fields (ordinary matter, magnetic field, gravitational field and vacuum energy), they have the same thermodynamic properties. Their energy densities  obey the same equation, which contains the corresponding matter densities and the pairs of the thermodynamically conjugate variables. In Minkowski vacuum, where the ordinary matter and magnetic and gravitational fields are absent, this thermodynamic approach automatically leads to zero cosmological constant.
  \end{abstract}
\pacs{
}

\maketitle

\tableofcontents
 
 
   \section{Introduction}
   
 In pre-geometric theories, gravitational field originates from the matter fields. For example, tetrads and metric fields may emerge as the non-linear combinations of the fermions fields, \cite{Akama1978,Wetterich2004,Wetterich2022,Diakonov2011,Diakonov2012,Volovik1990} or from the scalar fields.\cite{Wilczek1998,Addazi2025,Addazi2025b,Meluccio2025} That is why the gravitational field  also represents the matter in addition to the ordinary matter, which is represented by particles and fields. The same is in the Sakharov induced gravity, where metric becomes dynamical after integration over the fermionic and bosonic fields.\cite{Sakharov1968} In a similar way, in the Zel'dovich approach, the electrodynamic field becomes dynamical after integration over the charged particles.\cite{Zeldovich1967} The dynamical effective gravitational and electromagnetic fields emerge also in condensed matter with Weyl fermionic quasiparticles.\cite{Volovik2003} 
 
All this suggests that the curvature term in the Einstein–Hilbert action should be considered in the same way as the matter terms. In particular, the energy-momentum tensor of the gravitational field should be treated in the same way as the energy-momentum tensor of the ordinary matter, and thus it can be obtained by varying the curvature term in the Einstein–Hilbert action with respect to the spacetime metric $g^{\mu\nu}$. This leads to the appearance of a generalized energy-momentum tensors, which are applicable to all types of matter including gravitational field. According to Einstein's equations, the total generalized energy-momentum tensor, which includes the contributions of all matter fields including gravity, is equal to zero. Being zero it is automatically conserved, $\partial_\mu T^{\mu\nu}=\nabla_\mu T^{\mu\nu}=0$.
 
Here we apply this approach to the thermodynamics of the Universes that are homogeneous in space. This includes the de Sitter Universe, the Bonnor-Melvin-$\Lambda$ Universe and the static Einstein Universe. We consider these Universes as equilibrium thermodynamic states, ignoring their possible instability.

The plan of the paper is the following.

In Section \ref{GravFieldSection}, the gravitational field is considered as the specific matter field.
In this approach, the Einstein equations are obtained not  from the functional derivative of the action, $\delta S/\delta g^{\mu\nu}=0$, but from the conventional derivative of the Lagrangian, $\frac{d{\cal L}^{\rm M}}{d g^{\mu\nu}} =0$. This also allows to introduce the generalized energy-momentum tensor, which is applicable to any kind of matter including the gravitational field.  This approach looks somewhat similar to that developed by Chakraborty and Padmanabhan, \cite{Chakraborty2015a,Chakraborty2015b,Chakraborty2017} see also the recent paper \cite{Majhi2026}. But we do not use the connection between thermodynamics and horizon.

In Section \ref{dSsection}, this approach is applied to the thermodynamics of the de Sitter state. The de Sitter thermodynamics contains two components: the vacuum component (dark energy or cosmological constant) and the thermal gravitational component -- the Ricci scalar $R$. This component is characterized by temperature $T=H/\pi$, where $H$ is the Hubble parameter. The generalized energy density contains the pair of thermodynamically conjugate  variables, the Ricci scalar curvature $R$ and the gravitational coupling $K=1/16\pi G$. In the  $f(R)$ theory, $K=df/dR$ plays the role of the gravitational chemical potential.
 The thermal component of the de Sitter state has equation of state $w=1$, and thus looks as the Zel'dovich stiff matter. This gives rise to the de Sitter massless graviton discussed in Section \ref{MasslessGraviton}. 
 
 In Section \ref{BMsection}, the thermodynamics of the Universe with constant magnetic field is discussed -- the Bonnor-Melvin-$\Lambda$ Universe with cosmological constant. Here the generalized energy density has the same form as in the de Sitter thermodynamics, but the energy density of the de Sitter thermal component is replaced by the energy density of the magnetic field $\frac{B^2}{8\pi}$, while the $KR$ contribution remains the same. 

In Section \ref{StaticUniverse}, the thermodynamics of the $S^3$ static Einstein Universe is discussed. Again, the generalized energy density has the same form as in the de Sitter thermodynamics, with the de Sitter thermal component replaced by the energy density of the ordinary matter. 

In Section \ref{3formSec}, the dynamical dark energy is considered using the $3$-form gauge field introduced by Hawking. The generalized energy density has the same form, only another thermodynamic pair is added -- the curvature of the $3$-form field and its conjugate. The final result of the paper is in Eq.(\ref{GeneralMatter}). The generalized energy density of all three Universes contains the generalized Matter density, which includes all kinds of matter (ordinary matter, magnetic field, gravitational field and vacuum energy) and pairs of thermodynamically conjugate variables.
When ordinary matter, magnetic and gravitational fields are absent, and only vacuum energy remains, the result is a Minkowski vacuum with an automatically zero cosmological constant.

    \section{Gravitational field as specific matter field}
    \label{GravFieldSection}

We shall treat the gravitational field as any other field such as magnetic field, and thus the gravitational field is also the part of matter. This is natural in the approach, in which the gravitational field originates from the matter fields. The typical examples are the pre-geometric theories, in which the metric fields emerge as the non-linear combinations of the fermionic or bosonic fields, and the Sakharov theory, where the action for gravity is obtained by integration over the matter fields.

Let us start with the Einstein–Hilbert action written in the standard form:
 \begin{equation}
S=S^{\rm m}+S^{\rm grav}= S^{\rm m} - \int d^4x \sqrt{-g} KR \,.
\label{StandardAction}
\end{equation}
Here $S^{\rm m}$ is the action for the ordinary matter fields and $S^{\rm grav}$ is the gravitaional part of the action, where $R$ is the scalar curvature and $K=\frac{1}{16\pi G}$.

It is convenient to rewrite the Einstein–Hilbert action in the following way:
 \begin{equation}
S=-\int d^4x {\cal L}^{\rm M}  \,\,,\,\,  {\cal L}^{\rm M} ={\cal L}^{\rm m} +{\cal L}^{\rm grav}\,.
\label{Action2}
\end{equation}
Here ${\cal L}^{\rm m}$ is the Lagrangian of the ordinary matter; ${\cal L}^{\rm M}$ is the total matter Lagrangian, which includes the Lagrangian  ${\cal L}^{\rm m}$ of the ordinary matter and the Lagrangian ${\cal L}^{\rm grav}$ of the "matter of gravity" written using the Ricci tensor $R_{\mu\nu}$ instead of the Ricci scalar $R$:
 \begin{equation}
{\cal L}^{\rm grav}= \sqrt{-g} KR_{\mu\nu} g^{\mu\nu} \,.
\label{LagrangianGrav}
\end{equation}
With this form of the ${\cal L}^{\rm grav}$, the Einstein equations, which are usually obtained  from the functional derivative of the action, $\delta S/\delta g^{\mu\nu}=0$, can be obtained
 from the conventional derivative of the Lagrangian, $\frac{d{\cal L}^{\rm M}}{d g^{\mu\nu}} =0$.
 The reason is that the variation of the Riemann tensor $R_{\mu\nu}$ leads to the total derivative, and can be ignored with the proper choice of the boundary conditions. The same takes place for the matter action.

That is why the energy-momentum tensor of the gravitational field, as well as the energy-momentum tensor  of any other matter filed, can be obtained by the variation of the corresponding Lagrangian.
Then the total energy-momentum tensor of matter, including the "matter of gravity", is obtained from the derivative of the total Lagrangian with respect to the metric, $T^{\rm gen}_{\mu\nu}=\frac{2}{\sqrt{-g}} \frac{d{\cal L}^{\rm M}}{d g^{\mu\nu}}$.
According to the Einstein equations, this general energy momentum tensor equal to zero:
\begin{eqnarray}
T^{\rm gen}_{\mu\nu}=T^{\rm matter}_{\mu\nu}+ T^{\rm grav}_{\mu\nu}=\frac{2}{\sqrt{-g}} \frac{d{\cal L}^{\rm m}}{d g^{\mu\nu}}+\frac{2}{\sqrt{-g}} \frac{d{\cal L}^{\rm grav}}{d g^{\mu\nu}}  =
\nonumber
\\
=\frac{2}{\sqrt{-g}} \frac{d{\cal L}^{\rm M}}{d g^{\mu\nu}} =0
\,.
\label{ZeroGeneralized}
\end{eqnarray}
Being zero, the energy momentum tensor of the generalized matter field automatically obeys both the conventional and covariant conservation laws:
\begin{equation}
\partial_\mu T_{\rm gen}^{\mu\nu}=\nabla_\mu T_{\rm gen}^{\mu\nu} =0
\,.
\label{conservation}
\end{equation}
In this approach the concept of the energy-momentum pseudo-tensor\cite{LL2} is not needed.  This approach can be extended also to the Dirac field, see e.g. the review paper \cite{Shapiro2022}.
Note that the similar approach was applied by Stanyukovich in his book \cite{StanyukovichBook}, but only in the limit of weak gravitational field and in the frame of special relativity.
  
    \section{de Sitter Universe}
  \label{dSsection}
  
   \subsection{de Sitter solution}
    \label{dSsolutionSection}
    
The Einstein–Hilbert action for the de Sitter state contains two terms: the gravitational term with Riemann curvature and cosmological constant term (the vacuum energy):
 \begin{eqnarray}
S=-\int d^4x \sqrt{-g} (KR_{\mu\nu} g^{\mu\nu} +\epsilon_{\rm vac}) 
\,
\label{EinsteinAction}
\\
\epsilon_{\rm vac}=\frac{\Lambda}{8\pi G} \,\,,\,\, K=\frac{1}{16\pi G}\,.
\nonumber
\end{eqnarray}
Here we use the  signature $(-+++)$ for the Riemann tensor and for the metric, and $c=\hbar=1$.

As in the general case, the Einstein equations can be obtained from the conventional derivative of the Lagrangian in the  Einstein–Hilbert action with respect to the metric instead of the  functional derivative:
 \begin{equation}
\frac{d{\cal L}^{\rm M}}{d g^{\mu\nu}}=0=  (KR+\epsilon_{\rm vac}) g_{\mu\nu} -\frac{R_{\mu\nu}}{8\pi} \,.
\label{EinsteinEq2}
\end{equation}
or in the conventional form:
 \begin{equation}
 R_{\mu\nu}  -\frac{1}{2}Rg_{\mu\nu}-\Lambda g_{\mu\nu} =0\,.
\label{EinsteinEq}
\end{equation}
This equation is the same as that obtained by the variation of the action, $\delta S/\delta g^{\mu\nu}=0$.
  
The de Sitter solution of this equation is
 \begin{equation}
R_{\mu\nu} = -\Lambda g_{\mu\nu} \,\,,\,\,  R=-4\Lambda = -12 H^2 \,,
\label{dSEq}
\end{equation}
where $H$ is the Hubble parameter.
The de Sitter state is homogeneous and thus is non-compact. There are the problems related to the boundary, which can be cured by the introduction of the Gibbons-Hawking-York boundary term.\cite{York1972}
However, the boundary problem may concern the thermodynamics of the de Sitter state and also collective modes. These may depend on the non-dynamical terms. Such terms do not affect Einstein's equations, but they may contribute to the thermodynamics energy.

Eq.(\ref{ZeroGeneralized}) for the energy momentum tensor demonstrates that the equilibrium de Sitter state has zero generalized pressure, $P^{\rm gen}=0$, and zero generalized energy density, $\epsilon^{\rm gen}=0$:
\begin{equation}
\epsilon^{\rm gen}=P^{\rm gen}=0
\,.
\label{generalized}
\end{equation}
The equation $P^{\rm gen}=0$ has clear physical meaning: it is the total pressure, that is zero in the absence of the external pressure. Let us consider the physical meaning of the equation 
$\epsilon^{\rm gen}=0$. 

 \subsection{Two components of de Sitter thermodynamics}
    \label{TwoCompSection}

The key point is that  the de Sitter state represents the heat bath with the local temperature $T_{\rm dS}=H/\pi$.\cite{Volovik2026,Maxfield2022} It is temperature which is experienced by any object  immersed in the de Sitter environment, such as an hydrogen atom, which experiences the ionization with the rate determined by this temperature. This temperature is local, and is the same for all the co-moving observers. The local temperature suggests the local thermodynamics of the de Sitter state, which is described by the local entropy density $s_{\rm dS}$. The latter can be obtained in different ways, for example from the horizon entropy $S_{\rm hor}=\frac{A}{4G}$, where $A=4\pi/H^2$ is the area of the horizon. Dividing the horizon entropy by the volume $V_{\rm Hubble}=\frac{4\pi}{3H^3}$ bounded by the horizon, one obtains the entropy density of the de Sitter state:
 \begin{equation}
s_{\rm dS} =\frac{S_{\rm hor}}{V_{\rm Hubble}}=\frac{3H}{4G} =\frac{3\pi}{4G} T_{\rm dS} \,.
\label{EntropyDensity}
\end{equation}
The local entropy density of the de Sitter state, $s_{\rm dS}=\frac{3\pi}{4G} T_{\rm dS}$, can be also obtained directly using the local thermodynamics and the free energy density. This demonstrates that the local thermodynamics leads to the holographic bulk-horizon correspondence $s_{\rm dS}V_{\rm Hubble}=S_{\rm hor}$.

Eq.(\ref{EntropyDensity}) gives:
 \begin{equation}
T_{\rm dS} s_{\rm dS} =-KR\,,
\label{EntropyDensity2}
\end{equation}
which demonstrates that the thermal behaviour  of the de Sitter state is provided by the curvature $R$. 

This suggests the two-component thermodynamics of the de Sitter state:\cite{Volovik2025d} the vacuum component (dark energy) and the thermal component (gravity with its curvature). These are similar to the superfluid and normal components in the Landau two-fluid hydrodynamics. The vacuum component has equation of state  $w_{\rm vac}=-1$, i.e.
$P_{\rm vac}=  -\epsilon_{\rm vac}$. 
The equation of state for the gravitational component $P_n=w_n\epsilon_n$ can be found from the following considerations. First, the generalized pressure of the de Sitter state is the sum of the pressure from the "superfluid" (dark energy) component and the pressure from the "normal" (gravitational) component:
 \begin{equation}
P^{\rm gen}=P_{\rm vac}+ P_n=0\,.
\label{PressureTotal}
\end{equation}
This gives $P_n=-P_{\rm vac}=\epsilon_{\rm vac}$.
Secondly, the thermal component obeys the first law of thermodynamics:
\begin{equation}
\epsilon_n-T_{\rm dS}s_{\rm dS}=-P_n
\,.
\label{FirstLaw}
\end{equation}
Then from equations (\ref{PressureTotal}) and (\ref{EntropyDensity2}) one obtains the equation of state of the thermal component:
\begin{equation}
P_n=w_n \epsilon_n\,\,,\,\, w_n=1
\,.
\label{NormalEqState}
\end{equation}

The thermal component has the same equation of state, $w=1$, as the Zeldovich stiff matter,\cite{Zeldovich1962} and the vacuum and thermal components have equal energy densities, $\epsilon_n=\epsilon_{\rm vac}$.

 \subsection{Thermodynamically conjugate gravitational variables}
    \label{KRSection}

Now let us consider the meaning of the generalized energy density $\epsilon^{\rm gen}$ of the de Sitter state, which must be zero according to Eq.(\ref{generalized}). Two interpretations of $\epsilon^{\rm gen}$ can be suggested.

 1) The generalized energy density corresponds  the free energy density, and it is automatically zero:
 \begin{equation}
\epsilon^{\rm gen}\equiv F_{\rm dS}=\epsilon_{\rm vac}+ \epsilon_n -T_{\rm dS}s_{\rm dS}=0\,.
\label{FreeEnergy}
\end{equation}

2) The generalized energy density includes the thermodynamic conjugate variables $K$ and $R$, with $K$ playing the role of the chemical potential, see e.g. Ref. \cite{Volovik2025c}. Then one has:
 \begin{equation}
\epsilon^{\rm gen}\equiv \epsilon_{\rm vac}+ \epsilon_n + KR=0 \,.
\label{ChemicalPotential}
\end{equation}

For the de Sitter state, these two interpretations are equivalent, since according to Eq.(\ref{EntropyDensity2}) one has $T_{\rm dS}s_{\rm dS}=-KR$.
However, we shall show that for the magnetic Bonnor-Melvin universe with cosmological constant, only the second interpretation makes sense, see Section \ref{BMsection} and Eq.(\ref{KRgen}). This suggests that the second interpretation is general and applicable for all the homogeneous states of the Universe, including also the Einstein static Universe in Section \ref{StaticUniverse}, see Eq.(\ref{StaticUniverseKR}). The $KR$ term is the result of the Kronecker anomaly introduced by  Polyakov and Popov.\cite{PolyakovPopov2022} It emerges in the homogeneous systems due to jump of the number of the degrees of freedom at ${\bf k}=0$.

For the homogeneous state, in any coordinate system one may find the point where the metric is flat, $g^{\mu\nu} =\eta^{\mu\nu}$, and obtain $T^{\rm gen}_{\mu\nu}=d{\cal L}^{\rm M}/dh^{\mu\nu}$, where  $g^{\mu\nu} =\eta^{\mu\nu}+h^{\mu\nu}$ is the metric perturbation. We shall use this simplification for the Bonnor-Melvin-$\Lambda$ universe in the next section.

  \section{Bonnor-Melvin Universe with cosmological constant}
  \label{BMsection}

 \subsection{Homogeneous magnetic field}
  \label{MagneticFieldSection}

Homogeneous magnetic field also contributes to the Kronecker anomaly. In cosmology, homogeneous magnetic field is possible in the presence of the tuned cosmological constant, see e.g. \cite{Plebanski1979,Astorino2012,Zofka2019,Ahmed2024,Castro2024}. 
The Bonnor–Melvin-$\Lambda$ universe with homogeneous magnetic field has the fine-tuning between the magnetic field and cosmological constant (here we use $G=1$):
  \begin{equation}
F_{\mu\nu}F^{\mu\nu}=2\Lambda \,,
\label{HomogeneousF}
\end{equation}
 In this case there is no angular deficit, and the metric is
  \begin{eqnarray}
ds^2=-dt^2+dr^2 + \frac{1}{2\Lambda} \sin^2(\sqrt{2\Lambda}r)d\phi^2 + dz^2 \,,
\nonumber
\\
\sqrt{-g}=\frac{1}{\sqrt{2\Lambda}} \sin(\sqrt{2\Lambda}r) 
\,,
\label{HomogeneousEq}
\end{eqnarray}
and magnetic field is
  \begin{equation}
B(r)=\frac{1}{\sqrt{2}}  \sin(\sqrt{2\Lambda}r)  \,\,, \,\, F_{\mu\nu}F^{\mu\nu}=2B^2(r)g^{\phi\phi}(r)=2\Lambda
\,.
\label{HomogeneousH}
\end{equation}

The state is invariant under transformation $r\rightarrow r + 2\pi n/\sqrt{2\Lambda}$. But Eq.(\ref{HomogeneousF}) suggests the invariance under combined translations, as in the case of the pure de Sitter. For that, one should choose the proper coordinate system, which has no singularities.
Anyway, the real magnetic field is the field observed at point $r=0$, i.e. ${\bf B}=\sqrt{\Lambda}\hat{\bf z}$. The point $r=0$ can be arbitrarily chosen, which means the homogeneous magnetic field.  The energy density of magnetic field coincides with the vacuum energy density
 \begin{equation}
\epsilon_B=\frac{B^2}{8\pi}  = \frac{\Lambda}{8\pi}=\epsilon_{\rm vac} = \frac{3}{8\pi} H^2
\,,
\label{EnergyDensity}
\end{equation}
where $H$ is the de Sitter Hubble parameter in the pure de Sitter state.

 \subsection{Anisotropic energy-momentum tensor}
  \label{AnisotropicSection}

The other components of energy-momentum tensor of magnetic field are:
 \begin{equation}
T_{xx}=T_{yy}=-T_{zz}=T_{00}=\epsilon_B=\epsilon_{\rm vac} \,\,,\,\, T_{\mu\nu}g^{\mu\nu}=0
\,.
\label{EnergyMomentum}
\end{equation}
They provide the anisotropic pressure of magnetic field ${\bf B}$:
 \begin{eqnarray}
 T^\alpha_\beta(B)={\rm diag}( -\epsilon_B, P_{\perp B},P_{\perp B}, P_{\parallel B})=
 \nonumber
 \\
 = \epsilon_{\rm vac} \,{\rm diag}(-1,1,1,-1)\,.
\label{AnisotropicPressure}
\end{eqnarray}
This can be compared with  the isotropic pressure from the cosmological constant:
 \begin{equation}
T^\alpha_\beta(\Lambda)=  \epsilon_{\rm vac}\, {\rm diag}(-1,1,1,1)\,.
\label{IsotropicPressure}
\end{equation}

From Einstein equations
 \begin{equation}
R_{\mu\nu}- \frac{1}{2}Rg_{\mu\nu}= \Lambda g_{\mu\nu}-8\pi T_{\mu\nu}
\,.
\label{EinsteinEq}
\end{equation}
one obtains the curvature at $r=0$ and thus everywhere:
 \begin{equation}
R=-4\Lambda \,\,,\,\,R_{00}=R_{zz}=0 \,\,,  \,\, R_{xx}=R_{yy}=\frac{1}{2}R=-2\Lambda \,.
\label{R00}
\end{equation}
 
 One can also introduce the anisotropic pressure related to curvature:
  \begin{equation}
T^\alpha_\beta(R)=  \epsilon_{\rm vac}\, {\rm diag}(2,-2,-2,0)\,,
\label{CurvaturePressure}
\end{equation}
with $T^\alpha_\beta(B)+ T^\alpha_\beta(\Lambda) +T^\alpha_\beta(R)=0$:
  \begin{eqnarray}
T^\alpha_\beta(R)= \epsilon_{\rm vac}\, {\rm diag}(0,-2,-2,0) -  {\rm diag}(KR,0,0,0)\,.
\label{CurvaturePressureKR}
\end{eqnarray}

 \subsection{Generalized energy and conjugate gravitational variables}
  \label{ConjugateSection}
  
Equation $T^\alpha_\beta(B)+ T^\alpha_\beta(\Lambda) +T^\alpha_\beta(R)=0$ corresponds to the  generalized pressure, which is naturally zero, $P\equiv P^{\rm gen}=P_\perp=P_\parallel=0$.

 The last term on Eq.(\ref{CurvaturePressureKR}) is the same $KR$ term, which is responsible for Kronecker anomaly. This gives zero value for the generalized energy density:
  \begin{equation} 
  \epsilon^{\rm gen}=\epsilon_{\rm vac} +\epsilon_B+ KR =0 
  \,.
\label{KRgen}
\end{equation}
This equation is the Bonnor–Melvin-$\Lambda$ variant of the Eq.(\ref{ChemicalPotential}) for de Sitter state. The difference is that the temperature of the Bonnor–Melvin-$\Lambda$ state is zero, $T=0$, and thus  the normal component is absent. The thermal gravitational contribution $\epsilon_{\rm grav}$ in the de Sitter state  is substituted by the non-thermal magnetic contribution $\epsilon_B$ in the Bonnor–Melvin-$\Lambda$ state.  This is the fully equilibrium static state, where magnetic field compensates the vacuum pressure.  
    
 \section{Einstein static Universe}
 \label{StaticUniverse}

The Einstein static Universe is the elliptical $S^3$ Universe, which is homogeneous and isotropic, since all the points on the $S^3$ sphere are equivalent. The Einstein Universe has three components:\cite{Volovik2003,Volovik2024} the ordinary matter component, which plays the role of the normal component with equation of state $P_n=w_n\epsilon_n$; the space curvature component with equation of state $P_R=-\frac{1}{3}\epsilon_R$, where $\epsilon_R=KR$; and the vacuum component with equation of state $w_{\rm vac}=-1$.

The generalized pressure and generalized energy density are:
\begin{equation} 
  P^{\rm gen}=P_{\rm vac} +P_n+ P_R=0 
  \,,
\label{StaticUniversePressure}
\end{equation}
and
  \begin{equation} 
  \epsilon^{\rm gen}=\epsilon_{\rm vac} +\epsilon_n+ \epsilon_R=0 
  \,.
\label{StaticUniverseEnergy}
\end{equation}
Equations (\ref{StaticUniversePressure}) and Eq.(\ref{StaticUniverseEnergy}) determine the energy densities of matter $\epsilon_n$ and the dark energy density $\epsilon_{\rm vac}$:
 \begin{equation} 
\epsilon_n =-  \frac{2}{3(1+w_n)}KR \,\,,\,\,
\epsilon_{\rm vac} =-\left( 1- \frac{2}{3(1+w_n)}  \right) KR
  \,.
\label{EnergyDensities}
\end{equation}

Equation (\ref{StaticUniverseEnergy}) can be rewritten in the form of equations (\ref{ChemicalPotential}) and  (\ref{KRgen}) for de Sitter and Bonnor-Melvin universes:
 \begin{equation} 
  \epsilon^{\rm gen}=\epsilon_{\rm vac} +\epsilon_n+ KR=0 
  \,.
\label{StaticUniverseKR}
\end{equation}
For the radiation matter with $w=1/3$ one has $\epsilon_n=\epsilon_{\rm vac}$, which can be compared with $\epsilon_n=\epsilon_{\rm vac}$ in the de Sitter state and 
$\epsilon_B=\epsilon_{\rm vac}$ in the Bonnor–Melvin-$\Lambda$ Universe.

If one introduces the general Matter, which includes the ordinary matter in Einstein static Universe, magnetic field in the Bonnor–Melvin-$\Lambda$ Universe and gravitational field in de Sitter Universe, then equations  (\ref{ChemicalPotential}),  (\ref{KRgen}) and (\ref{StaticUniverseKR})  can be written in the general form:
 \begin{equation} 
  \epsilon^{\rm gen}=\epsilon_{\rm vac} +\epsilon_{\rm Matter}+ KR=0 
  \,,
\label{Matter}
\end{equation}
In all three homogeneous universes, the generalized energy density contains the pair of thermodynamic variables $R$ and $K$. In the  $f(R)$ theory, $K=df/dR$.

In the Section \ref{3formSec} we consider the emergence of the other thermodynamic pairs. They are responsible for the vacuum energy density making $\epsilon_{\rm vac}$ another part of general Matter.

\section{$3$-form field and cosmological constant problem}
\label{3formSec}

\subsection{Dynamic vacuum energy from Hawking $3$-form gauge field}
\label{DynamicVacSec}

Till now we considered the constant vacuum energy $\epsilon_{\rm vac}=\frac{\Lambda}{8\pi}$, i.e. the cosmological constant $\Lambda$ (here we use $G=1$).
One of the ways to make the vacuum energy dynamical is to use the $3$-form gauge field introduced by Hawking.\cite{Hawking1984,Duff1989,Wu2008} In this approach developed in Refs.\cite{KlinkhamerVolovik2008,KlinkhamerVolovik2008a} another pair of thermodynamically conjugate variables emerges: the $4$-form curvature $q$ and the corresponding analogue of the chemical potential $\mu=d\epsilon_{\rm vac}/dq$.  From the Einstein equations and equation for the $3$-form field it follows  that the generalized energy density has now the form:
 \begin{equation} 
  \epsilon^{\rm gen}=\epsilon_{\rm vac}(q) - \mu q  +\epsilon_{\rm Matter}+ KR=0 
  \,.
\label{qfield}
\end{equation}
The first two terms in the the right-hand side of Eq.(\ref{qfield}) enter the Einstein equations in the form of the cosmological constant:
 \begin{equation} 
  \epsilon_{\rm vac}(q) - \mu q \equiv \frac{\Lambda}{8\pi}
  \,.
\label{CC}
\end{equation}

In the same way as the gravitational and electromagnetic fields, the $q$-field is also the matter field. That is why its energy density $\epsilon_{\rm vac}(q)$ must be also included into the generalized Matter energy density. Then Eq.(\ref{qfield}) can be written in the general form
 \begin{equation} 
  \epsilon^{\rm gen}=  \epsilon_{\rm Matter}+ KR  - \sum_a \mu^{(a)} q^{(a)}=0
  \,,
\label{GeneralMatter}
\end{equation}
where $ \epsilon_{\rm Matter}$ includes all types of matter (ordinary matter,  dark energy, the "matter of gravity" and all types of the gauge fields).
It also contains the pairs of the thermodynamically conjugate variables, where 
 we also took into account the possibility of several different $3$-form fields, see Eq.(10) in Ref. \cite{KlinkhamerVolovik2008c}. The same situation takes place in the multi-component liquid with densities $n^{(a)}$, where in the absence of the external pressure and at zero temperature one has $\epsilon - \sum_a \mu^{(a)} n^{(a)}=0$.

\subsection{Minkowski vacuum and cosmological constant problem}
\label{MinkowskiSec}

In the Minkowski vacuum, only the vacuum variables are left and Eq.(\ref{qfield}) becomes
 \begin{equation} 
  \epsilon^{\rm gen}=\epsilon_{\rm vac}(q) - \mu q =\frac{\Lambda}{8\pi}=0 
  \,.
\label{qfieldMinkowski}
\end{equation}
The thermodynamic nullification of the cosmological constant $\Lambda$ provides the natural solution of the cosmological constant problem. In the Minkowski vacuum one has $\Lambda=0$. The huge Planck-scale energy density $\epsilon_{\rm vac}(q)$ is naturally compensated by the thermodynamic "counterterm" $\mu q$ without any fine-tuning.
This also follows from the first law of thermodynamics: at $T=0$ one has
$\epsilon_{\rm vac}(q) - \mu q =-P_{\rm vac}$, and in the absence of external pressure, $P_{\rm vac}= P_{\rm external}=0$, one obtains $\Lambda=0$. This thermodynamic property does not depend on the microscopic structure of the system, i.e. on the ultraviolet trans-Planckian physics.
Thermodynamics is a fundamental theory that operates in the infrared limit and is independent of the ultraviolet cutoff.

The de Sitter state, even for very large values of $\Lambda$, is unstable due to its local temperature, which leads to the thermal creation of ordinary matter. Then de Sitter state eventually decays towards the Minkowski state with $\Lambda \rightarrow 0$. The rate of decay of the de Sitter state remains an unsolved problem, see e.g. Ref.\cite{Polyakov2018}.
The phenomenological approach to this problem gives the power law decay of the vacuum energy, see Eq.(34) in Ref. \cite{Volovik2026}. The power law decay is in agreement with the results obtained using several different approaches. \cite{Padmanabhan2003,Padmanabhan2005,KlinkhamerVolovik2016,Markkanen2018,Markkanen2018a,Roman2020,Gong2021}

\section{Massless graviton in de Sitter as second sound in two-fluid hydrodynamics}
\label{MasslessGraviton}

Thermodynamics is the universal phenomenon, that is applicable to different systems including condensed matter. Condensed matter experiences the same thermodynamic compensation of the energy density. For example, in superfluid $^4$He at $T=0$ and in the absence of external pressure one has $\epsilon -\mu n= -P=0$, where $n$ and $\mu$ are the thermodynamically conjugate variables:  the density of $^4$He atoms and the chemical potential corresondingly.

 There is another interesting connection to the thermodynamics of condensed matter systems -- the analogy with the Landau two-fluid hydrodynamics. 
 In the Bonnor–Melvin-$\Lambda$ Universe and in de Sitter Universe, matter represents the thermal components with different equations of state: arbitrary $w_n$ for ordinary thermal matter in Bonnor–Melvin-$\Lambda$ Universe and $w_n=1$ for the gravitational component  in de Sitter Universe. The latter has the same equation of state as the Zeldovich stiff matter.\cite{Zeldovich1962}  While in ordinary matter the speed of sound is $s^2=w_n c^2$,  the speed of sound in Zeldovich stiff matter is equal to the speed of light. Since in the de Sitter state the normal component is of the gravity component, this sound corresponds to the de Sitter graviton.

It appears that this graviton can be also considered as the second sound in the two-fluid thermodynamics of de Sitter state.\cite{Volovik2026a} Let us consider first the second sound in the two-fluid hydrodynamics of superfluid $^4$He. The velocity $s_2$ of the second sound in the two-fluid hydrodynamics has the following general form, see e.g. \cite{Schmitt2014}:
\begin{equation}
s_2^2=\frac{TS^2}{C_V\rho} \,\frac{\rho_s}{\rho_n} \,.
\label{EntropyRatio}
\end{equation}
Here $T$ is the temperature of the liquid helium; $S$ is the entropy density of the liquid, which is concentrated in the normal component of the liquid; $C_V$ is the specific heat; $\rho_s$ and $\rho_n$ are densities of superfluid and normal components correspondingly; and $\rho =\rho_n +\rho_s$ is the liquid density.

The same equation for the speed of second sound is applicable to the de Sitter two-fluid thermodynamics. With the de Sitter local thermodynamic variables $T=T_{\rm dS}$; $S=C_V=s_{\rm dS}$; $\rho_s=\epsilon_{\rm vac}/c^2$;
$\rho_n=\rho_s$,
one obtains that the velocity of the second sound in Eq.(\ref{EntropyRatio}) coincides with the speed of light:
\begin{equation}
 s_2=c \,.
\label{SpeedSecondSound}
\end{equation}
It also coincides with the speed of sound in Zeldovich stiff matter, since the second sound is the propagating mode in the thermal gravitational component, which behaves as Zeldovich stiff matter.

At first glance this seems strange: why would Landau's classical two-fluid hydrodynamic equations lead to a relativistic propagation at the speed of light. But apparently the two-fluid approach is quite general. It follows from the general laws of thermodynamics, which apply also to the relativistic quantum vacuum. It is the local thermodynamics of the de Sitter state with its local temperature $T_{\rm dS}=H/\pi$ which is responsible for the relativistic character of this mode. The relativistic spectrum cannot be obtained if instead of the local temperature one uses  the Gibbons-Hawking temperature $T_{GH}=H/2\pi$ of the cosmological horizon. That is why the equation (\ref{SpeedSecondSound}) provides the additional support to the local de Sitter thermodynamics.
The local de Sitter thermodynamics allows also to study the problem of stability of  the regular black holes with de Sitter core.\cite{Lobo2026,Volovik2026Sing}

Although the two fluid thermodynamics of the de Sitter state fully determines the spectrum of  the massless de Sitter graviton, it would be interesting to derive this spectrum directly from the de Sitter dynamics. However,  there is still no consensus regarding the modes in the de Sitter background, see e.g. \cite{Garidi2003,Akhmedov2017,Gazeau2023,Sadekov2024,Hinterbichler2025,Glavan2025,Rajantie2025,Todorov2026}. 
To avoid these uncertainties, the gravitational Kronecker anomaly must be properly taken into account.

   \section{Conclusion}

The thermodynamics of the homogeneous Universes is constructed, considering the gravitational field as a part of matter fields. It is demonstrated that the de Sitter Universe, the Bonnor-Melvin-$\Lambda$ Universe and the static Einstein Universe, although they have different types of matter fields (ordinary matter, magnetic field, gravitational field and vacuum energy), they have the same thermodynamic properties.  Their energy densities  obey the same equation (\ref{GeneralMatter}), which contains the generalized matter density and the pairs of the thermodynamically conjugate variables. In the absence of the ordinary matter, magnetic and gravitational fields, the Universe represents the Minkowski vacuum with automatically zero cosmological constant.

\end{document}